**Experimental and Theoretical Studies of the N($^2$D) + H$_2$ and D$_2$ Reactions**

Dianailys Nuñez-Reyes,[a,b] Cédric Bray,[a,b] Kevin M. Hickson,[a,b] Pascal Larrégaray,[a,b] Laurent Bonnet[a,b] and Tomás González-Lezana[c]

[a]*Université de Bordeaux, Institut des Sciences Moléculaires, F-33400 Talence, France*
[b]*CNRS, Institut des Sciences Moléculaires, F-33400 Talence, France*
[c]*Instituto de Física Fundamental, CSIC, IFF-CSIC Serrano 123, 28006 Madrid*

**TOC entry text**

This study reports the first kinetic measurements of the N($^2$D) + H$_2$, D$_2$ reactions below 200 K.

**Abstract**

This study reports the results of an experimental and theoretical investigation of the N($^2$D) + H$_2$ and N($^2$D) + D$_2$ reactions at room temperature and below. On the experimental side, a supersonic flow (Laval nozzle) reactor was employed to measure rate constants for these processes at temperatures as low as 127 K. N($^2$D) was produced indirectly by pulsed laser photolysis and these atoms were detected directly by pulsed laser induced fluorescence in the vacuum ultraviolet wavelength region. On the theoretical side, two different approaches were used to calculate rate constants for these reactions; a statistical quantum mechanical (SQM) method and a quasi-classical trajectory capture model including a semi-classical correction for tunneling (SC-Capture). This work is described in the context of previous studies, while the discrepancies between both experiment and theory, as well as between the theoretical results themselves are discussed.

**Introduction**

The gas-phase reactions of electronically excited atoms C($^1$D), N($^2$D), O($^1$D) and S($^1$D) with hydrogen in particular have been extensively studied by both experimental[1] and theoretical[2] means. While these processes share certain common features, there are also important differences which affect the properties of individual systems. All four reactions are characterized by the presence of deep wells over the potential energy surfaces (PES) connecting reagents with products, corresponding to an XH$_2$ intermediate species. Similarly, all four reactions are exothermic, where the exothermicities of the C($^1$D) and S($^1$D) reactions are considerably smaller than those of the corresponding N($^2$D) and O($^1$D) ones. While the C($^1$D), O($^1$D) and S($^1$D) + H$_2$ reactions are fast, barrierless processes, the presence of a small barrier in the entrance channel of the N($^2$D) + H$_2$ reaction plays a crucial role in the kinetics and dynamics of this system.

The dynamics of the N($^2$D) + H$_2$, D$_2$ reactions have been the subject of several earlier experimental investigations. Dodd et al.[3] used a pulsed electron beam to dissociate N$_2$ in the presence of H$_2$ in their work, detecting the infrared emission from vibrationally excited NH (X$^3\Sigma^-$) radicals produced by the N($^2$D) + H$_2$ reaction. They measured inverted nascent vibrational distributions for the NH product, a result that was interpreted as an indication of a direct abstraction type mechanism. Later work by Umemoto and coworkers on the N($^2$D) + H$_2$, D$_2$ reactions[4, 5] used the pulsed two-photon dissociation of NO at 275.3 nm to produce N($^2$D) atoms while NH (ND) radicals were detected by laser induced fluorescence (LIF). They determined cooler nascent NH (ND) vibrational distributions than those obtained by Dodd et al.,[3] corresponding to reaction occurring mostly through an insertion type process. These results were confirmed by crossed molecular beam (CMB) experiments employing a radiofrequency discharge of N$_2$ for N($^2$D) production coupled with tunable electron impact ionization and a time of flight (TOF) mass spectrometer with a quadrupole mass filter for mass-selective product detection.[6, 7] Centre-of-mass product angular distributions determined for the *m/e* = 15 NH products were found to be nearly backward–forward symmetric, a characteristic sign of a long-lived intermediate complex resulting from an insertion type mechanism.

Theoretical investigations [2, 8, 9] on the PES developed by Pederson et al.[8] for the 1 $^2A''$ ground electronic state revealed that the dynamics of the title reactions correspond indeed to an insertion mechanism. In particular, the application of a statistical quantum mechanical (SQM)[10] and semi-classical[11] statistical methods produced both rotational and vibrational distributions[2] in good agreement with experimental[4, 5] and quantum mechanical (QM) results. These same statistical approaches also successfully reproduced product translational energy distributions, laboratory angular distributions and TOF spectra obtained in CMB experiments.[7] The marked forward-backward scattering symmetry displayed by differential cross sections (DCS) is also interpreted as a clear signal of the complex-forming character of the reaction, and therefore, unsurprisingly, the statistical predictions reproduce nicely the corresponding state-to-state angular distributions obtained with exact QM methods.[7] It should also be noted that a semi-classical[11] statistical treatment only accounting for the long range isotropic forces predicted state distributions with good accuracy. Using an improved version of the PES of Pederson et al.[8] developed by Ho et al.,[12] cross sections and rate constants for the reactions N($^2$D) + H$_2$/D$_2$/HD were calculated[13] by means of the SQM approach and quasi-classical trajectories (QCT) calculations in an attempt to investigate both the rotation and isotope effect on these processes. The theoretical predictions for the excitation function of the N($^2$D)+D$_2$ reaction exhibited some deviation from measurements by Liu[14] as the energy increases beyond the threshold, but according to spectroscopic studies[15] the real transition from an insertion mechanism (with a barrier of ≈ 0.08 eV) to a mechanism dominated by abstraction dynamics (with a slightly larger linear barrier of ≈ 0.21 eV) is expected at larger energies. Lin et al.[16] analysed in some detail the possible contribution of the abstraction pathways in the overall dynamics of the N($^2$D) + H$_2$ reaction, calculating the collision time distributions of the NH$_2$ intermediate complex and some specific state-to-state DCSs. Despite the contribution to the backward scattering peak for the angular distribution of the N($^2$D) + H$_2$ → H + NH(v$_f$ = 3) process, both the rotational distributions and DCSs obtained by means of QM and QCT calculations[16] are those expected from a reaction following complex-forming dynamics. Moreover, DCSs obtained by means of time dependent QM methods for N($^2$D) + H$_2$[17] and N($^2$D) + D$_2$[18] on a different PES[19] at collision energies up to 0.3 and 0.2 eV respectively, still display the characteristic forward-backward symmetry of such insertion mechanisms when the calculation is performed only on the ground electronic state and no Renner-Teller effects are included. Therefore, in this work, we have extended the SQM calculations performed in Bañares et al.[13] to cover a lower temperature regime explored by the new experiments reported here.

On the kinetics side, there are several previous measurements of the rate constants for the N($^2$D) + H$_2$ reaction at room temperature using a range of experimental techniques[20-27] with values within the range (1.7 - 5.0) × 10$^{-12}$ cm$^3$ s$^{-1}$. Umemoto et al.[27] also measured the room temperature rate constant for the N($^2$D) + D$_2$ reaction with a value equal to 1.4 × 10$^{-12}$ cm$^3$ s$^{-1}$, in excellent agreement with the earlier result of Suzuki et al.[28] To date, the only temperature dependent investigation of the N($^2$D) + H$_2$ reaction has been performed by Suzuki et al.[28] who measured the rate constants for the N($^2$D) + H$_2$ and D$_2$ reactions over the 213 – 300 K temperature range. These experiments employed a pulsed electron beam to dissociate N$_2$ in the presence of H$_2$, coupled with the direct detection of N($^2$D) atoms at 149 nm by a resonance absorption method. They found that the rate constants for both reactions decreased at lower temperature, deriving Arrhenius expressions equal to $k(T) = 4.6 \times 10^{-11}$ exp(-880/$T$) cm$^3$ s$^{-1}$ and $k(T) = 3.9 \times 10^{-11}$ exp(-970/$T$) cm$^3$ s$^{-1}$ for the N($^2$D) + H$_2$ and N($^2$D) + D$_2$ reactions respectively.

Apart from the fundamental interest surrounding the N($^2$D) + H$_2$ reaction, this process is also expected to play a role in the chemistry of planetary atmospheres. In particular, Titan's atmosphere is characterized by low temperatures (70-200 K), and is predominantly composed of molecular nitrogen (~95%) with trace amounts of other gases including molecular hydrogen (~0.1%). Here, N$_2$ photolysis in the vacuum ultraviolet wavelength range in the upper atmosphere leads to the production of significant quantities of N($^2$D) atoms, while the non-reactive removal of N($^2$D) through the N($^2$D) + N$_2$ quenching reaction is thought to be very slow at low temperature.[29] Previous photochemical models[30] have already identified the N($^2$D) + H$_2$ → NH + H reaction as an important source of NH radicals in Titan's atmosphere, that contributes to the formation of methanimine (CH$_2$NH) through the NH + CH$_3$ → CH$_2$NH + H reaction. Nevertheless, the currently recommended rate constants for the N($^2$D) + H$_2$ reaction, based on the review by Herron,[29] have never been measured over the appropriate temperature range.

**Experimental Methods**

A continuous supersonic flow reactor as described in previous work[31, 32] was used to perform the present kinetic measurements. This apparatus has been modified progressively to allow various atomic reagents in their ground (C($^3$P),[33-35] H($^2$S)[34-39] and D($^2$S)[39-41]) and excited (N($^2$D)[42-44] and O($^1$D)[38, 45-47]) electronic states to be produced and detected. Only a single Laval nozzle was used for the present investigation, employing the carrier gases nitrogen and argon, both of which quench N($^2$D) atoms inefficiently.[28, 48] This allowed temperatures of 177 K (N$_2$) and 127 K (Ar) to be attained. The specific flow characteristics for these temperatures are described in earlier work.[42] For experiments conducted at 296 K, the Laval nozzle was removed and slow flow conditions were employed in the reactor.

As photolytic sources of N($^2$D) atoms are quite rare, these atoms were generated indirectly instead during the present study. Here we employed the fast reaction between ground state atomic carbon C($^3$P) and nitric oxide (NO),[49] which is known to produce N($^2$D) atoms with a substantial yield[50, 51]

$$C(^3P) + NO \rightarrow N(^2D, {}^4S) + CO \quad (1a)$$
$$\rightarrow O(^3P) + CN \quad (1b)$$

as demonstrated in recent work.[42-44] Here, C($^3$P) atoms were produced photolytically along the entire length of the supersonic flow by the 266 nm pulsed photodissociation of carbon tetrabromide (CBr$_4$) precursor molecules. Previous work[33] has shown that excited state C($^1$D) atoms are also produced with a 10-15% yield with respect to C($^3$P) atoms. CBr$_4$ molecules were injected into the gas flow upstream of the nozzle by passing a small fraction of the carrier gas over solid CBr$_4$, held in a flask that was maintained at a known pressure and temperature, leading to estimated concentrations of less than $3.2 \times 10^{13}$ cm$^{-3}$ within the supersonic flow. Coreagent NO was mixed with the carrier gas flow before passing into the Laval nozzle reservoir, with concentrations used in the range $(2.7-6.2) \times 10^{14}$ cm$^{-3}$ within the supersonic flow.

N($^2$D) atoms were detected directly during this work by pulsed laser induced fluorescence at 116.745 nm (VUV LIF). More details of the third harmonic generation method used to produce intense tunable narrowband radiation around this wavelength can be found in our earlier work.[42] The VUV beam was steered through a sidearm containing baffles which reduced the level of scattered light from the VUV excitation source and residual UV radiation within the reactor. The sidearm was constantly flushed with argon or nitrogen to limit the absorption of the VUV laser by residual gases such as NO or by CBr$_4$ vapour. Fluorescence emission from excited N($^2$D) atoms within the supersonic flow was collected orthogonally to both the supersonic flow and the VUV excitation beam by a solar blind photomultiplier tube

(PMT). The PMT and its associated collection optics were isolated from the reactor by a lithium fluoride (LiF) window. Additionally, the volume between the LiF window and PMT was evacuated to prevent secondary emission losses through absorption by atmospheric $O_2$. A LiF lens was placed between the LiF window and the PMT to focus the fluorescence emission onto the PMT photocathode. The PMT output signal was fed into an amplifier before integration by a boxcar system. Under these conditions, the amplified VUV LIF signal could only be recorded after 15 microseconds had elapsed following the photolysis laser pulse due to amplifier saturation issues. A digital delay generator allowed the pulsed photolysis and probe lasers to be synchronized as well as the boxcar integration system and oscilloscope used to monitor the fluorescence signal. Each time point was an average of 30 individual laser shots, while at least 70 laser shots were recorded for each decay profile. At least 15 time points were recorded with the probe laser firing prior to the photolysis laser to establish the baseline level. Gases (Linde Ar 99.999%, Xe 99.999%, $D_2$ 99.8%, Air Liquide $H_2$ 99.9999%, $N_2$ 99.999%, NO 99.9%) were used without further purification directly from the cylinders. Calibrated mass-flow controllers were used to regulate gas flows into the reactor.

**Theoretical Methods**

**Statistical Quantum Mechanics**

As mentioned in the introduction, we have extended the SQM calculations reported in Bañares et al.[13] to cover a lower temperature regime. Details about the method have been given before and can be, for instance, found in that previous work on the title reaction.[13] Here it suffices to remember that the state-to-state reaction probability is approximated as follows:[52, 10]

$$P^J_{vjl,v'j'l'}(E) = \left|S^J_{vjl,v'j'l'}(E)\right|^2 = \frac{p^J_{vjl}(E) p^J_{v'j'l'}(E)}{\sum_{v''j''l''} p^J_{v''j''l''}(E)} \qquad (2)$$

$p_{vjl}^J(E)$ is the capture (or complex-formation) probability, i.e., the probability to enter the well when coming with the total energy $E$, the total angular momentum $J$, the orbital angular momentum $l$ and with H$_2$/D$_2$ in the initial rovibrational state ($v, j$). $p_{v'j'l'}^J(E)$ is the analogous quantity for the products NH/ND in the final state ($v', j'$). Accordingly, the reagent state-resolved integral cross section (ICS) is then calculated as:

$$\sigma_{vj}(E) = \frac{\pi}{k^2(2j+1)} \sum_{Jl}(2J+1) \sum_{v'j'l'} P_{vjl,v'j'l'}^J(E) \qquad (3)$$

with $k^2 = 2\mu E/\hbar^2$, $\mu$ being the atom-diatom reduced mass. Here, the zero energy has been taken at the rovibrational initial state of the reagents, thus making the collision energy equal to $E$. The state-resolved rate constant is obtained as follows:

$$k_{vj}(T) = g\sqrt{\frac{8\beta^3}{\pi\mu}} \int_0^\infty \sigma_{vj}(E) E e^{-\beta E} dE \qquad (4)$$

where $g$ is the electronic degeneracy (equal to 1/5 in this case) and $\beta = (k_B T)^{-1}$. The PES employed in the study is the same as before, that by Ho et al.[12] but the value of the asymptotic distance reached for the log-derivative propagation, performed as in Bañares et al.[13] within the centrifugal sudden approximation to calculate the capture probabilities for Equation (2), have been extended up to $\approx$ 27 Å and $\approx$ 37 Å for the reactants and products arrangements, respectively. The calculations of the ICSs have been performed up to a 0.5 eV collision energy including all H$_2$ ($v, j$) and D$_2$ ($v, j$) rovibrational states below $E_{max}$ = 0.9 eV or 0.95 eV, depending on the cases, reaching a maximum value of the total angular momentum $J_{max}$ = 44 and 46, respectively.

**Semi-Classical Capture Model**

QCT calculations, including correction for tunneling, have also been performed to estimate the rate constants. Since the reaction is highly exothermic and there is no exit-barrier, the probability that a trajectory having entered the intermediate well recrosses the entrance-barrier towards the reagents is very low. In other words, the capture probability is a good approximation to the reaction probability. The reagent state-resolved capture rate constant $k_{vj}(T)$ can thus be computed using equations (3) and (4), but replacing $\sum_{v'j'l'} P^J_{vjl,v'j'l'}(E)$ by $p^J_{vjl}(E)$.

When running $N_{traj}$ trajectories from initial conditions corresponding to $E$ and the reagent quantum state ($J, v, j, l$) (see further below for their definition), the capture probability $p^J_{vjl}(E)$ is given by

$$p^J_{vjl}(E) = \sum_{n=1}^{N_{traj}} p_n / N_{traj} \qquad (5)$$

where $p_n$ is the capture probability assigned to the $n^{th}$ trajectory. For a reaction with no entrance barrier, this probability is 1 if the trajectory enters the well, 0 otherwise. But the process under scrutiny involves an entrance barrier of about 0.08 eV requiring to account for tunneling in the capture dynamics through a semiclassical treatment. In the rest of this section, we first define the initial conditions, and second present our semiclassical approach of tunneling.

The initial conditions of the molecular system are selected as follows. The radial distance $R$ between the N atom and the center-of-mass of $H_2$ ($D_2$) is taken at 15 Å and the momentum $P$ conjugate to $R$ is given by $P = -[2\mu E-(l/R)^2]^{1/2}$ (this momentum is negative since the system is incoming). The phase space state is fully specified by $R$, $P$ and the ten action-angle coordinates $J, M, l, j, n, \alpha, \beta, \alpha_l, \alpha_j$ and $q$. $J, l$ and $j$ are the classical total, orbital and rotational angular momenta in $\hbar$ unit. $M$ is the projection of $J$ on the $Z$ axis of the laboratory frame. $n$ is the vibrational action, or classical analog of the vibrational quantum number. $n$ is taken to be equal to $v$. The five angles are conjugate to the five previous actions. The transformation from these coordinates to Cartesian coordinates within which trajectories are run is fully detailed in Appendix A of Bonnet et al.[53] (Note that $J, j, l, j_{xy}$ and $l_{xy}$ cannot be strictly 0 in the expression of appendix A. $\alpha$ and $\beta$ must then be chosen accordingly, e.g. $\alpha =$

π/4, β= π/2.) Whenever necessary, a back transformation can be performed at any instant of the run in order to recover $R$, $P$ and the action-angle coordinates.

Tunneling through the entrance barrier is taken into account as follows; the classical Hamiltonian for the molecular system reads:

$$H = \frac{P^2}{2\mu} + \frac{l^2}{2\mu R^2} + \frac{j^2}{2mr^2} + \frac{p^2}{2m} + V(R,r,\gamma) \quad (6)$$

where $m$ is the $H_2$ reduced mass and $\gamma$ the Jacobi angle between $R$ and the vector $r$ between the two H or D atoms. If for a given trajectory, capture is classically allowed, *i.e.* if the trajectory overcomes the entrance barrier, $p_n$ is taken to be one. In practice, each trajectory which reaches $R_{min}= 3a_0$ is considered to be captured (the top of the minimum $C_{2v}$ energy barrier lies at approx. $R=4a_0$). If capture is classically prohibited, a turning point along the $R$ coordinate shows up at a given instant $t$. As a consequence, capture may still happen by tunneling through the barrier, the probability of which is computed in the following way. The trajectory is stopped at the turning point where the radial momentum $P=0$ and $l_t$, $j_t$, $p_t$, $R_t$, $r_t$ and $\gamma_t$ are respectively the values of $l$, $j$, $p$, $R$, $r$ and $\gamma$. From this point inward, it is assumed that the vibration motion can be decoupled from the ro-translational one. At the turning point, the potential energy thus reads:

$$V(R_t,r_t,\gamma_t) = V_t(R_t,r_{eq},\gamma_t) + v_t(r_t) \quad (7)$$

where $v_t(r_t)$ and $V_t(R_t,r_{eq},\gamma_t)$ are, respectively, the potentials for the diatom on one hand, and the remaining degrees-of-freedom on the other hand. $r_{eq}$, the equilibrium diatom distance is found to be almost constant from the reactant to the barrier top. At the turning point, the diatom vibrational energy can thus be approximated by

$$E_{vib}^t = \frac{p_t^2}{2m} + v_t(r_t) \quad (8)$$

and the $H_2$ rotational kinetic energy is

$$E_R^{K,t} = \frac{j_t^2}{2mr_t^2}. \quad (9)$$

As a consequence, the effective collision energy reads:

$$E^t = E + E_{rovib} - E^t_{vib} - E^{K,t}_R \qquad (10)$$

where we recall that $E$ is the initial collision energy, and $E_{rovib}$ is the initial rovibrational energy. Assuming that the rotation and vibration motions are frozen during barrier crossing, tunneling is computed along the radial coordinate $R$, at collision energy $E^t$, through the effective potential

$$V^t_{eff}(R, r_{eq}, \gamma_t) = \frac{l_t^2}{2\mu R^2} + V_t(R, r_{eq}, \gamma_t) \qquad (11)$$

The tunneling probability for the n$^{th}$ trajectory, $p_n$, is estimated using the Wentzel-Kramers-Brillouin (WKB) semi-classical model[54, 55] by

$$p_n = \frac{1}{1 + e^{2\theta(E^t, l_t)}} \qquad (12)$$

where the phase integral $\theta(E^t, L_t)$ reads

$$\theta(E^t, L_t) = \int_{R_{min}}^{R_t} \sqrt{2\mu \left(V^t_{eff}(R, r_{eq}, \gamma_t) - E^t\right)/\hbar^2} \, dR \qquad (13)$$

This model is called the Semi-Classical Capture model in the following (SC-Capture).

To compute the rate constants, the capture reagent state-resolved cross-sections $\sigma_{v=0,j=0,1,2}(E)$ have been computed from 10 to 300 meV collision energies, for each 5 meV below 200 meV and for each 10 meV beyond. At each collision energy, the maximum value of the orbital angular momentum, $l_M$, has been estimated in the following way. The tunneling probability along the reaction path ($\gamma$=90°) has been computed for increasing values of $l$ (eq. 12 and 13), until it was found to be a thousand times lower than the overall tunneling probability at $l=0$. More than 2000 trajectories have been run for each value of $l$. Convergence has been checked. The experimental rotational distribution has been accounted for to compare with measured rate constants.

**Results and Discussion**

**Experimental Rate Constants and Comparison with Theory**

As the coreagents NO and $H_2$ ($D_2$) were present in excess with respect to the minor reagents $N(^2D)$ and $C(^3P)$ for all experiments reported here, the pseudo-first-order approximation could be applied. Consequently, the $N(^2D)$ concentration, $[N(^2D)]$, which was considered to be proportional to its VUV LIF signal, $I_{N(^2D)}(t)$, was expected to vary as a function of time according to the expression

$$I_{N(^2D)}(t) = A(\exp(-k'_a t) - \exp(-k'_b t)) \qquad (14)$$

where $k'_a$ is the pseudo-first-order rate constant for $N(^2D)$ loss, $k'_b$ is the pseudo-first-order rate constant for $N(^2D)$ formation and $t$ is time. $A$ is a constant that represents the theoretical maximum $N(^2D)$ fluorescence signal in the absence of secondary losses (that is, when $k'_a = 0$). Under the present experimental conditions, several processes contribute to $k'_a$ including the reactions of $N(^2D)$ with NO ($k_{N(^2D)+NO}[NO]$), $X_2$ ($k_{N(^2D)+X_2}[X_2]$, where X = H or D), precursor molecule $CBr_4$ ($k_{N(^2D)+CBr_4}[CBr_4]$) as well as diffusion ($k_{N(^2D),\text{diff}}$). As $N(^2D)$ atoms are produced by reaction (1a), their rate of formation, $k'_b$, is dependent on the pseudo-first-order rate constant for this process, $k_{C(^3P)+NO}[NO]$, in addition to any other removal processes for $C(^3P)$ atoms such as reaction with $CBr_4$, $k_{C(^3P)+CBr_4}[CBr_4]$, and diffusion $k_{C(^3P),\text{diff}}$. In any event, as the first 15 microseconds of all temporal profiles recorded during these experiments could not be exploited due to amplifier saturation brought about by scattered light from the photolysis laser pulse, a fitting procedure employing a single exponential function of the form

$$I_{N(^2D)}(t) = A \times \exp(-k'_a t) \qquad (15)$$

was preferred instead. Two such profiles, recorded at room temperature for the reaction of $N(^2D)$ with $D_2$ are displayed in Figure 1.

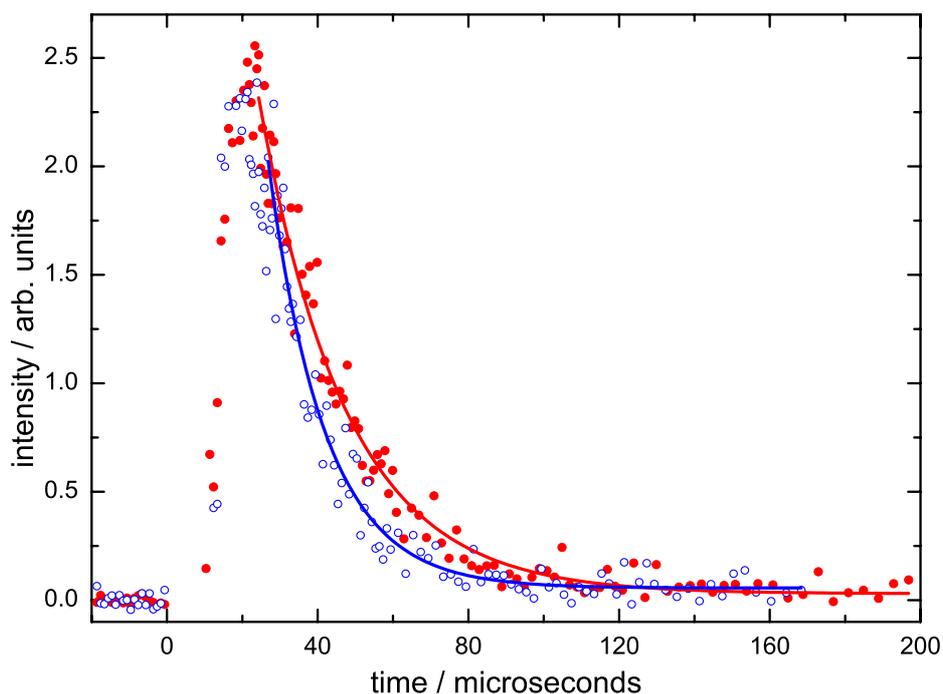

**Figure 1** N($^2$D) fluorescence intensity as a function of time, recorded at 296 K. (Open blue circles) [D$_2$] = 1.2 × 10$^{16}$ cm$^{-3}$; (solid red circles) without D$_2$. [NO] = 6.2 × 10$^{14}$ cm$^{-3}$ for this series of experiments. Solid blue and red lines represent single exponential fits to the data allowing $k'_a$ to be derived.

In this case, it was necessary to check that N($^2$D) formation was close enough to completion to ensure that the datapoints included in the fit were well described by expression (3). In practice, this could be easily verified by checking for deviations from linearity on plots of $\ln(I_{N(^2D)})$ versus $t$.

In contrast to our recent investigation of the N($^2$D) + C$_2$H$_2$ reaction,[44] which was hampered by interference from the competing C($^3$P) + C$_2$H$_2$ reaction (leading to fewer available C($^3$P) atoms for N($^2$D) production through reaction (1a)), the reaction C($^3$P) + H$_2$ → CH + H($^2$S) is highly endothermic (+95.4 kJ mol$^{-1}$) so that it can be neglected at room temperature and below. Consequently, the majority of other potential secondary reactions that might interfere with the kinetics of the N($^2$D) + X$_2$ reactions have already been discussed in our recent work on the N($^2$D) + NO[42] and N($^2$D) + C$_x$H$_{2x+2}$[43] reactions. Other possible additional secondary reactions that could occur in the present study are those of the NX product of the N($^2$D) + X$_2$ reaction with X$_2$, NO and CBr$_4$. The NH + H$_2$ reaction is endothermic and characterized by a large activation barrier and should therefore be slow at low temperature. Mullen & Smith[56] studied the NH + NO reaction down to 53 K, measuring an increase in the reaction rate from as the temperature falls, while Yamasaki et al.[57] determined that the major products should be OH + N$_2$. In this case, it is not expected that subsequent reactions of the OH product will alter the N($^2$D) kinetics. No information could be found regarding the NH + CBr$_4$ reaction although it is unlikely that this process will lead to N($^2$D) formation.

For each temperature, kinetic decays similar to those shown in Figure 1 were recorded for at least 7 different coreagent X$_2$ concentrations, with at least two individual decays at any single X$_2$ concentration. As [NO] and [CBr$_4$] were maintained at constant values for any single series of experiments, the observed differences between derived $k'_a$ values was only due to the changing value of [X$_2$]. In this respect, a linear dependence of the derived $k'_a$ values was observed as a function of the coreagent X$_2$ concentration. A linear least-squares fit to these data yielded the second-order rate constant, $k_a$, at a given temperature. The second-order plots for the N($^2$D) + H$_2$ reaction at 296 K (solid red circles), 177 K (solid green triangles) and 127 K (solid blue squares) are displayed in Figure 2 along with their respective best-fit lines.

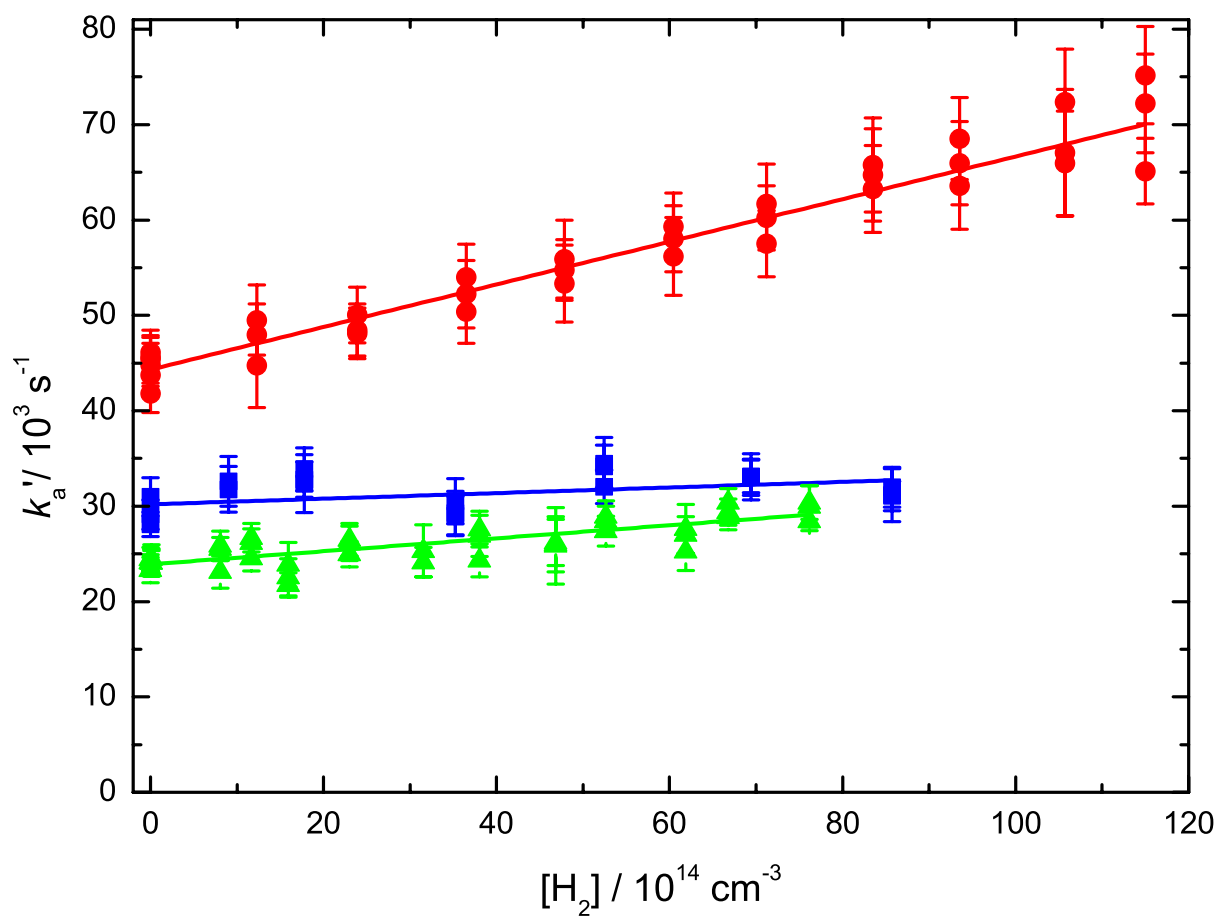

**Figure 2** Variation of the measured pseudo-first-order rate constant $k_a'$ as a function of [$H_2$]. (Solid red circles) 296 K; (solid green triangles) 177 K (solid blue squares) 127 K. Solid red, green and blue lines represent weighted linear least-squares fits to the data. The error bars (1σ) are derived from exponential fits such as those displayed in Figure 1.

The temperature dependences of the experimental rate constants for the N($^2$D) + H$_2$ and N($^2$D) + D$_2$ reactions are shown in Figure 3 alongside earlier work over a range of temperatures. These values are also compared with those derived by the present SQM and SC-Capture calculations. The measured and calculated second-order rate constants are listed in Tables 1 and 2 respectively. Table 3 lists the previous experimental kinetics studies of these reactions at room temperature.

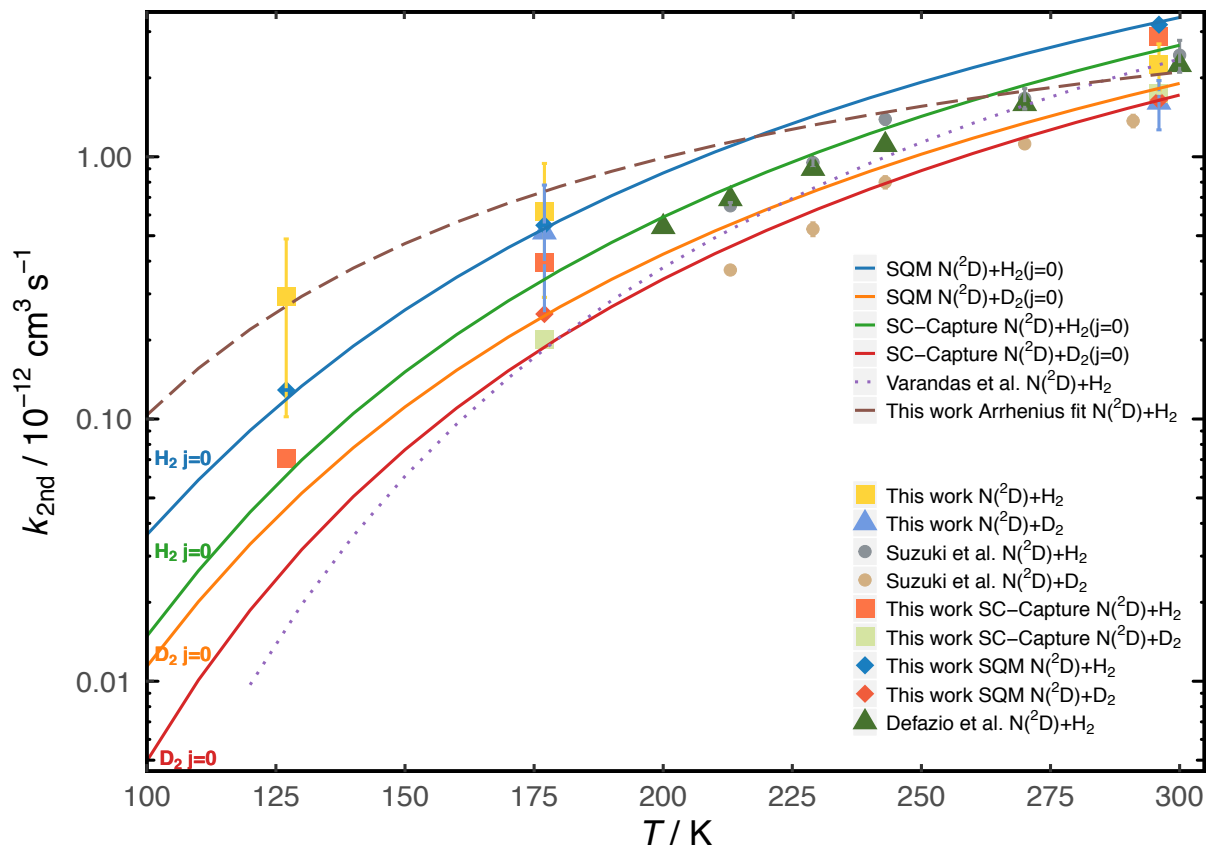

**Figure 3** Rate constants for the N($^2$D) + X$_2$ reactions (X = H or D) as a function of temperature. **The N($^2$D) + H$_2$ reaction** : (Solid grey circles) Suzuki et al.;[28] (solid yellow squares) this work - experiment; (solid orange squares) this work – SC-Capture; (solid blue diamonds) this work - SQM theory; (solid green triangles) Defazio & Petrongolo.[58]; (dotted purple line) Varandas et al.[59]; (dashed brown line) Arrhenius fit to the present N($^2$D) + H$_2$ rate constants. **The N($^2$D) + D$_2$ reaction** : (solid light brown circles) Suzuki et al.;[28] (solid light blue triangles) this work - experiment; (solid light green squares) this work – SC-Capture; (solid red diamonds) this work - SQM theory. Error bars on the present measurements represent the combined statistical (1σ) and systematic errors. Theoretical rate constants consider the non-equilibrium low temperature rotational distributions of H$_2$ and D$_2$. Solid lines represent the theoretical rate constants for individual rotational levels of H$_2$ (j = 0) and D$_2$ (j = 0) derived by both SQM and SC-Capture.

**Table 1** Measured second-order rate constants for the N($^2$D) + X$_2$ reactions

| T / K | $N^b$ | [H$_2$] / 10$^{15}$ cm$^{-3}$ | $k_{N(^2D)+H_2}$ / 10$^{-13}$ cm$^3$ s$^{-1}$ | $N^b$ | [D$_2$] / 10$^{14}$ cm$^{-3}$ | $k_{N(^2D)+D_2}$ / 10$^{-13}$ cm$^3$ s$^{-1}$ |
|---|---|---|---|---|---|---|
| 296 | 36 | 0 – 11.50 | (22.35 ± 4.55)$^c$ | 33 | 0 – 11.53 | (16.10 ± 3.43)$^c$ |
| 177 ± 2[a] | 38 | 0 - 7.61 | (6.17 ± 3.26) | 32 | 0 - 7.43 | (5.15 ± 2.65) |
| 127 ± 2 | 20 | 0 - 8.57 | (2.94 ± 1.92) | | | |

[a]Uncertainties on the calculated temperatures represent the statistical (1σ) errors obtained from Pitot tube measurements of the impact pressure. [b]Number of individual measurements. [c]Uncertainties on the measured rate constants represent the combined statistical (1σ) and estimated systematic errors. The systematic errors are considered to be 20 % of the nominal rate constant at 296 K and 50 % at lower temperature (see text for more details).

**Table 2** Calculated second-order rate constants for the $N(^2D) + X_2$ reactions

| T / K | $k_{N(^2D)+H_2}$(SQM) / $10^{-13}$ cm$^3$ s$^{-1}$ | $k_{N(^2D)+H_2}$ (SC-Capture) / $10^{-13}$ cm$^3$ s$^{-1}$ | $k_{N(^2D)+D_2}$(SQM) / $10^{-13}$ cm$^3$ s$^{-1}$ | $k_{N(^2D)+D_2}$ (SC-Capture) / $10^{-13}$ cm$^3$ s$^{-1}$ |
|---|---|---|---|---|
| 296 | 31.9 | 28.8 | 18.3 | 17.5 |
| 177 | 5.48 | 3.95 | 2.51 | 2.01 |
| 127 | 1.29 | 0.71 | | |

**Table 3** Previous room temperature experimental data

| T / K | $k_{N(^2D)+H_2}$ / $10^{-13}$ cm$^3$ s$^{-1}$ | $k_{N(^2D)+D_2}$ / $10^{-13}$ cm$^3$ s$^{-1}$ | Reference |
|---|---|---|---|
| 300 | 22.8 ± 2.3 | | Umemoto et al.[27] |
| 300 | 35.0 ± 10.0 | | Fell et al.[24] |
| 300 | 21.0 ± 3.0 | | Husain et al.[22] |
| 300 | 17.0 ± 5.0 | | Husain et al.[21] |
| 300 | 27.0 ± 2.0 | | Black et al.[23] |
| 300 | 50.0 | | Black et al.[20] |
| 300 | 18.0 ± 8.0 | | Whitefield and Hovis[25] |

| | | |
|---|---|---|
| 300 | 23.0 ± 5.0 | Piper et al.[26] |
| 300 | 24.4 ± 3.4 | Suzuki et al.[28] |
| 291 | 13.7 ± 0.7 | Suzuki et al.[28] |
| 295 | 13.7 ± 1.9 | Umemoto et al.[27] |

The experimental rate constants for both the N($^2$D) + H$_2$ and N($^2$D) + D$_2$ reactions ($k_{N(^2D)+H_2}$(296 K) = 2.2 × 10$^{-12}$ cm$^3$ s$^{-1}$ and $k_{N(^2D)+D_2}$(296 K) = 1.6 × 10$^{-12}$ cm$^3$ s$^{-1}$) are seen to be in excellent agreement with the previously measured values at room temperature ($k_{N(^2D)+H_2}$(300 K) = (1.7 – 3.5) × 10$^{-12}$ cm$^3$ s$^{-1}$, excluding the early work of Black et al.[20] and $k_{N(^2D)+D_2}$(300 K) = 1.4 × 10$^{-12}$ cm$^3$ s$^{-1}$). The measured rate constants for both reactions decrease below room temperature to less than 10$^{-12}$ cm$^3$ s$^{-1}$; values which are considered to be at the limit of measurable reactivity using the Laval nozzle technique. A similar problem was encountered in our recent study of the N($^2$D) + C$_x$H$_{2x+2}$ reactions,[43] where a more detailed discussion of this issue can be found. It should be noted that due to inefficient gas-phase spin conversion of both H$_2$ and D$_2$ between their ortho (o) and para (p) forms, the room temperature population ratios of these gases (H$_2$ o/p (296 K) = 3, D$_2$ o/p (296 K) = 2) are preserved at lower temperature. While the pseudo-first-order rate constants measured here for the N($^2$D) + H$_2$ reaction display a clear progression with [H$_2$] at 296 K and even at 177 K (see Figure 2), the data obtained at 127 K present very little variation as a function of [H$_2$]. To account for the significantly larger error of these low temperature values, a systematic uncertainty of 50 % has been attributed to all the rate constant measurements below 296 K (a systematic uncertainty of 20 % was used at 296 K). Despite the large uncertainty on the present low temperature rate constants, it can be seen from Figure 3 that a similar trend to the previous work of Suzuki et al.[28] over the 213-300 K range is observed for the N($^2$D) + H$_2$ reaction. An Arrhenius type fit ($k = A$ exp($-E_a / RT$) ) to the present data for the N($^2$D) + H$_2$ reaction yields the parameters $A =$

$9.5 \times 10^{-12}$ cm$^3$ s$^{-1}$ and $E_a/R = 452$ K, which compare with the currently recommended values of $A = 4.6 \times 10^{-11}$ cm$^3$ s$^{-1}$ and $E_a / R = 880$ K from Herron.[29] At 150 K (a representative temperature for Titan's atmosphere) the recommended Arrhenius parameters yield a rate constant $k_{N(^2D)+H_2}(150\ K) = 1.3 \times 10^{-13}$ cm$^3$ s$^{-1}$ which is almost four times smaller than the value derived in this work of $k_{N(^2D)+H_2}(150\ K) = 4.7 \times 10^{-13}$ cm$^3$ s$^{-1}$.

The calculated rate constants derived by SC-Capture and SQM are also displayed in Figure 3. These values take into consideration the non-equilibrium rotational distributions of H$_2$ and D$_2$ below room temperature that arise because of inefficient spin-conversion in the experiments, although the differences between these values and those obtained at equilibrium is expected to be small. The present SC-Capture calculations give a rate constant at 177 K ($k_{N(^2D)+H_2}(177\ K) = 3.95 \times 10^{-13}$ cm$^3$ s$^{-1}$) that is in good agreement with the experimental value although the value derived at 127 K ($k_{N(^2D)+H_2}(127\ K) = 0.71 \times 10^{-13}$ cm$^3$ s$^{-1}$) is lower and outside of the experimental error bars. For the SQM calculations, rate constants of $k_{N(^2D)+H_2}(177\ K)$ and $k_{N(^2D)+H_2}(127\ K)$ of $5.5 \times 10^{-13}$ cm$^3$ s$^{-1}$ and $1.3 \times 10^{-13}$ cm$^3$ s$^{-1}$ respectively are obtained; values that are well within the uncertainties of the present measurements. The theoretical investigation performed by Bañares et al.[13] reported rate constants for the title reactions between 200 K and 400 K. The QCT and SQM values at 296 K (interpolated from Fig. 5 of their paper) for the N($^2$D) + H$_2$ reaction are about $21.6 \times 10^{-13}$ cm$^3$ s$^{-1}$ and $30 \times 10^{-13}$ cm$^3$ s$^{-1}$, respectively.

In the case of the N($^2$D) + D$_2$ reaction, it was only possible to extend the kinetic measurements down to 177 K, as this reaction was found to be too slow to measure at even lower temperatures. Consequently, as experimental rate constants were only recorded at two temperatures, it is inappropriate to provide Arrhenius parameters in this instance. The derived rate constant at 177 K ($k_{N(^2D)+D_2}(177\ K) = (5.2 \pm 2.7) \times 10^{-13}$ cm$^3$ s$^{-1}$) is somewhat larger than the value predicted by Suzuki et al.[28] of $k_{N(^2D)+D_2}(177\text{ K}) = 1.6 \times 10^{-13}$ cm$^3$ s$^{-1}$ based on Arrhenius parameters ($A = 3.9 \times 10^{-11}$ cm$^3$ s$^{-1}$ and $E_a / R = 970$ K) derived from their data measured above 213 K. The large value determined by the present measurement almost certainly reflects the experimental difficulties of measuring small rate constants with the CRESU technique. Nevertheless, the present SQM calculations yield a value for $k_{N(^2D)+D_2}(177\text{ K})$ of $2.51 \times 10^{-13}$ cm$^3$ s$^{-1}$ which remains within the error bars of the current measurements. The SC-capture result of $k_{N(^2D)+D_2}(177\text{ K}) = 2.01 \times 10^{-13}$ cm$^3$ s$^{-1}$ is somewhat lower than both the experimental and SQM results at this temperature.

Results for the N($^2$D) + D$_2$ reaction at 296 K in the calculation by Bañares et al.[13] are also slightly smaller than the present ones, since their QCT calculation yielded a value of $14.7 \times 10^{-13}$ cm$^3$ s$^{-1}$ whereas their previous statistical calculation produced a rate constant of $16.9 \times 10^{-13}$ cm$^3$ s$^{-1}$.

The comparison between our theoretical results reveals that the SQM method leads to larger rate constants than those obtained by means of the SC-Capture calculation for both N($^2$D) + H$_2$ and N($^2$D) + D$_2$ reactions, but relative differences seem to decrease as the temperature increases. SC-Capture calculations ignore possible back-scattering after entering the intermediate well, as well as quantum reflection[54, 55] for the classically captured trajectories. Both effects, if relevant could lower reaction probability and hence rate constants. The fact that SC-Capture calculations predict lower rate constants than SQM thus suggests tunneling through the entrance barrier is lower in the former theoretical model. The effect of including tunneling in the capture model is examined in the supplementary information file. In addition, the possible origin of the quantitative differences between rate constants obtained by the SQM and SC-Capture methods is discussed. As an example, for N($^2$D) + H$_2$, although the inclusion of tunneling in the classical capture model significantly increases reactivity ( 2.94 times at 127K), tunneling is greater in SQM, leading to a rate constant 1.82 times greater than that of the SC-Capture model at 127 K (5.35 times greater than the value given by classical capture).

Consistent with previous investigations of the title reactions using the SQM approach,[13] no special dependence on the initial rotational state of either $H_2$ or $D_2$ is observed. On the contrary, SC-Capture rate constants for the reaction initiated with reactants in their ground states, $H_2$ (j=0) and $D_2$ (j=0), are smaller than those with rotationally excited $H_2$ (j=1) and $D_2$ (j=1), especially for $N(^2D) + H_2$.

**Conclusions**

This work describes a joint experimental and theoretical study of the gas-phase $N(^2D) + H_2$ and $N(^2D) + D_2$ reactions at low temperature. Experimentally, a supersonic flow apparatus was used, coupled with the pulsed laser photolysis method and pulsed laser induced fluorescence for the production and detection of $N(^2D)$ atoms respectively. Rate constants for the $N(^2D) + H_2$ reaction were measured between 127 K and 296 K, whereas rate constants were obtained for the $N(^2D) + D_2$ reaction between 177 K and 296 K. Theoretical estimates for the rate constants of these reactions were calculated over the same temperature range using two different approaches; a statistical quantum method and quasi-classical trajectory calculations. Good agreement is obtained between both theoretical methods and experiment at room temperatures, although at lower temperature the statistical predictions are seen to provide a better comparison with the measured rate constants than the quasi-classical trajectory ones. Such differences between the two sets of calculations might be ascribed to quantitative differences for tunneling through the entrance barrier, which need now to be accurately investigated.

**Conflicts of Interest**

There are no conflicts of interest to declare

**Acknowledgements**

KMH acknowledges support from the French program ''Physique et Chimie du Milieu Interstellaire'' (PCMI) of the CNRS/INSU with the INC/INP co-funded by the CEA and CNES as well as funding from the ''Program National de Planétologie'' (PNP) of the CNRS/INSU. CB acknowledges support from the project OSCAR ANR-15-CE29-0017 of the French National Research Agency (ANR) through the provision of a postdoctoral fellowship. TGL acknowledges support from Spanish MICINN with Project FIS2017-83157-P.